**Nazarovets, S.A.**
Kyiv National University of Culture and Arts,
14, Chyhorina St., Kyiv, 01042, Ukraine,
+380 44 285 7412, serhii.nazarovets@gmail.com

# Black Open Access in Ukraine: Analysis of Downloading Sci-Hub Publications by Ukrainian Internet Users

***Introduction.*** High subscription fees to scholarly research journals provoke researchers to use illegal channels of access to scientific information. Analysis of statistical data on downloads of scholarly research papers by Ukrainian Internet users from illegal web resource can help to define gaps in information provision at the institutional or the state level for each scientific field.

***Problem Statement.*** To conduct an analysis of behavior and geography of downloads of scholarly research publications from illegal web resource Sci-Hub by Ukrainian Internet users within the period from September 1, 2015 to February 29, 2016.

***Purpose.*** To assess the information needs of Ukrainian researchers who download scientific papers from Sci-Hub.

***Materials and Methods.*** The used file is available at public domain and contains complete data of downloads of scholarly research articles from Sci-Hub for the period from September 1, 2015, to February 29, 2016. Queries of users with Ukrainian IP-addresses have been selected. Using DOI of downloaded articles enables finding the publishers and journal brands with the help of CrossRef API, whereas using the All Science Journal Classification (ASJC) codes makes it possible to identify the subject.

***Results.*** The study has shown that the most documents downloaded related to natural sciences (primarily, chemistry, physics, and astronomy), with Elsevier publications being the most frequently quered by Ukrainian users of Sci-Hub and Internet users from Kyiv downloading the papers most actively.

***Conclusion.*** The obtained data are important for understanding the information needs of Ukrainian researchers and can be used to formulate an optimal subscription policy for providing access to information resources at Ukrainian R&D institutions.

*K e y w o r d s:* scholarly research journals, usage statistics, Open Access, Sci-Hub, and Ukraine.

At the end of the last century, the entire scholarly research community was expecting that the rapid development of electronic publishing accompanied by a significant reduction in the costs associated with the production, distribution, and storage of books and journals would lead to a significant drop in prices for scholarly research periodicals. However, notwithstanding with appearance of new effective channels for the dissemination of scientific information, the prices of leading world-class publishers of academic literature have not dropped, which hinders a scientific progress and is accompanied with aggravating information inequalities between scholarly researchers from different countries. In response to this, the world scientific community has united for the global open access (unrestricted and permanent access to scientific documents through the Internet, free of charge, without limitations imposed by licenses and copyright infringement) and proposed several functional models to provide open access to scholarly research publications. For a long while, were Golden Open Access that provides the publication of results in the open access journal, and Green Open Access that encourages researchers to publish preprints of their research papers on open specialized web resources have been deemed to be the most promising models [1].

Despite optimistic forecasts and efforts of Open Access advocates, it has been implemented was much slower than expected. As of 2017, only 35—50% of all relevant scholarly research publications can be found on Web resources that support Gold Open Access or the Green Open Access Model [2]. As a result, new channels of access to scientific publications, which do not require subscriptions, payments, and bureaucracy have appeared. Among them, there are ResearchGate academic social network, #icanhazpdf hashtag in Twitter, Sci-Hub and LibGen websites with pirated copies. Such initiatives are called differently: Rogue Open Access, Robin Hood Open Access, Guerilla Open Access, Bibliogifts [3]. Herein after, they are referred to as «Black Open Access».

One of the most popular illegal resource base of scholarly research publications is Sci-Hub (https://sci-hub.la) that aims at removing all barriers for as wide as possible dissemination of knowledge in the modern society. Sci-Hub is a script that downloads HTML and PDF pages from the Internet and saves them on its servers. These publications are available to users for further free download upon input of URL, DOI or title of required document. Upon the results of this research, Sci-Hub can provide an instant access to 2/3 of all scholarly research papers or more than 85% of all scholarly works published in subscription journals. For large popular publishers like Elsevier, this index is even higher, as over 97% of journal articles are illegally stored on Sci-Hub servers and available for free download [4].

For a certain time, Black Open Access was believed to be supported mainly by researchers from poor and developing countries. However, according to John Bohannon's analysis of user queries (28 million queries for 6 months) Sci-Hub is widely used worldwide, in various countries and fields of science. In particular, American and European researchers also actively use Sci-Hub [5, 6]. Upon the poll results, about 37% of those who got a pirated copy via Sci-Hub noted that they could do this using the legal access as well; 23% of respondents told they intentionally chose Sci-Hub because of disagreement with price policy of large academic publishers, while 17% said the access to full versions of scholarly research articles via Sci-Hub was easier than that via other legal channels [7].

John Bohannon and Alexandra Elbakian, the founder of Sci-Hub, have placed statistical data of user queries in Sci-Hub for the period from September 1, 2015, to February 29, 2016, on the open Internet (https://doi.org/10.5061/dryad.q447c), which allowed all interested researchers from many countries to analyze the use of Sci-Hub in their countries [8–10]. This information has provoked numerous debates on ethics, economic models, and the future of scientific communications [11—13]. The statistics on downloads from Sci-Hub are given in *.csv format that contains data on the date of query, user's geographical coordinates, country, city, and DOI.

This information gave a reason for studying information needs of Ukrainian researchers by analyzing behavior and geography of Ukrainian users of Sci-Hub. The results are given below.

Only queries from Ukrainian IP-address were selected from the initial file containing full information on downloads from Sci-Hub for the period from September 1, 2015 to February 29, 2016. The data on DOI of downloaded documents have enabled to identify publishers and journal using API CrossRef and the journal topics using All Science Journal Classification (ASJC) codes.

The obtained results show that during the reporting period users from Kyiv downloaded the largest number of documents from Sci-Hub, with a significant gap between the number of downloads from the capital of Ukraine and that from other regions reported. The number of downloads from Sci-Hub for users from various oblast centers differs, but this difference is not as important as in the previous case (see Table).

**Statistical Data on Downloads from Sci-Hub from Ukraine**

| Oblast | Population in 2016*, thousand | Urban population in 2016*, thousand | Downloads from Sci-Hub | Downloads to urban population ratio |
|---|---|---|---|---|
| Kyiv City | 2906569 | 2906569 | 186838 | 0,064 |
| Kharkiv Oblast | 2718616 | 2193536 | 36657 | 0,017 |
| Lviv Oblast | 2534174 | 1544862 | 28337 | 0,018 |
| Donetsk Oblast | 4265145 | 3870115 | 25154 | 0,006 |
| Vinnytsia Oblast | 1602163 | 813387 | 21248 | 0,026 |
| Dnipropetrovsk Oblast | 3254884 | 2722102 | 11470 | 0,004 |
| Odesa Oblast | 2390289 | 1597346 | 8554 | 0,005 |
| Chernivtsi Oblast | 909893 | 391810 | 7745 | 0,02 |
| Chernihiv Oblast | 1044975 | 675292 | 4376 | 0,006 |
| Sumy Oblast | 1113256 | 763606 | 3237 | 0,004 |
| Zaporizhizhia Oblast | 1753642 | 1353773 | 2866 | 0,002 |
| Cherkasy Oblast | 1242965 | 706205 | 1908 | 0,003 |
| Kyiv Oblast | 1732235 | 1078011 | 1275 | 0,001 |
| Zakarpattia Oblast | 1259158 | 466948 | 1153 | 0,002 |
| Ternopil Oblast | 1065709 | 475236 | 1144 | 0,002 |
| Mykolaiv Oblast | 1158207 | 790634 | 868 | 0,001 |
| Ivano-Frankivsk Oblast | 1382352 | 604509 | 847 | 0,001 |
| Kherson Oblast | 1062356 | 650474 | 787 | 0,001 |
| Luhansk Oblast | 2205389 | 1916176 | 666 | 0,0003 |
| Poltava Oblast | 1438948 | 891724 | 569 | 0,0006 |
| Zhytomyr Oblast | 1247549 | 733667 | 547 | 0,0007 |
| Volhynian Oblast | 1042668 | 545553 | 437 | 0,0008 |
| Khmelnytskyi Oblast | 1294413 | 729835 | 287 | 0,0004 |
| Rivne Oblast | 1161811 | 552980 | 241 | 0,0004 |
| Kirovohrad Oblast | 973150 | 611667 | 46 | 0,0001 |

\* Ukraine's population as of January 1, 2017; Kyiv, the State Statistical Service of Ukraine, 2017, p. 83, URL: http://database.ukrcensus.gov.ua/PXWEB2007/ukr/publ_new1/index.asp

Assuming that in oblasts with larger population, especially, with larger urban population, there is a higher probability of higher statistics of downloads from Sci-Hub the Table shows downloads to urban population ratio. Hence, the highest ratios belong to Internet users from Kyiv City, Vinnytsia, Chernivtsi, Lviv, and Kharkiv Oblasts. For more accurate comparison of statistics of downloads, it would be advisable to take into consideration the number of employees of R&D and educational establishments and students in various oblasts of Ukraine, however, for the time being, these statistical data are not available.

The statistics of downloads from Sci-Hub by users from Donetsk Oblast significantly exceeds the number of queries from majority of oblasts of Ukraine, even despite the fact that since April 2014, in Donbas, there has been warfare as a result of Russian aggression against Ukraine, and a large part of R&D and higher education establishments shifted to other territories. For instance, one of the leading HEEs of Ukraine, the Donetsk National University together with the major part of its students and teachers moved to Vinnytsia. Presumably, for this reason Vinnytsia Oblast is the second, next to Kyiv City, in terms of downloads to urban population ratio. Thus, in peacetime, the statistics of queries to Sci-Hub for the users from Donetsk and Luhansk Oblasts might be much higher.

The general statistics of queries to Sci-Hub do not have any information on downloads by users from the Autonomous Republic of the Crimea and Sevastopol. However, it is unlikely that the users from this region of Ukraine have never queried Sci-Hub. The absence of information on downloads can be explained by the use of programs and services for anonymous access to the Internet after the occupation.

In addition to the Sci-Hub user geographic structure, the statistical data contain unique digital identifiers of downloaded documents — DOI. This enables to understand which publishers, subjects, and journals are the most popular among Ukrainian Sci-Hub users. They are Elsevier (105 168 downloads), Springer Nature (46 354 downloads), American Chemical Society (40 728 downloads), and Wiley-Blackwell (28 567 downloads) (see Figure).

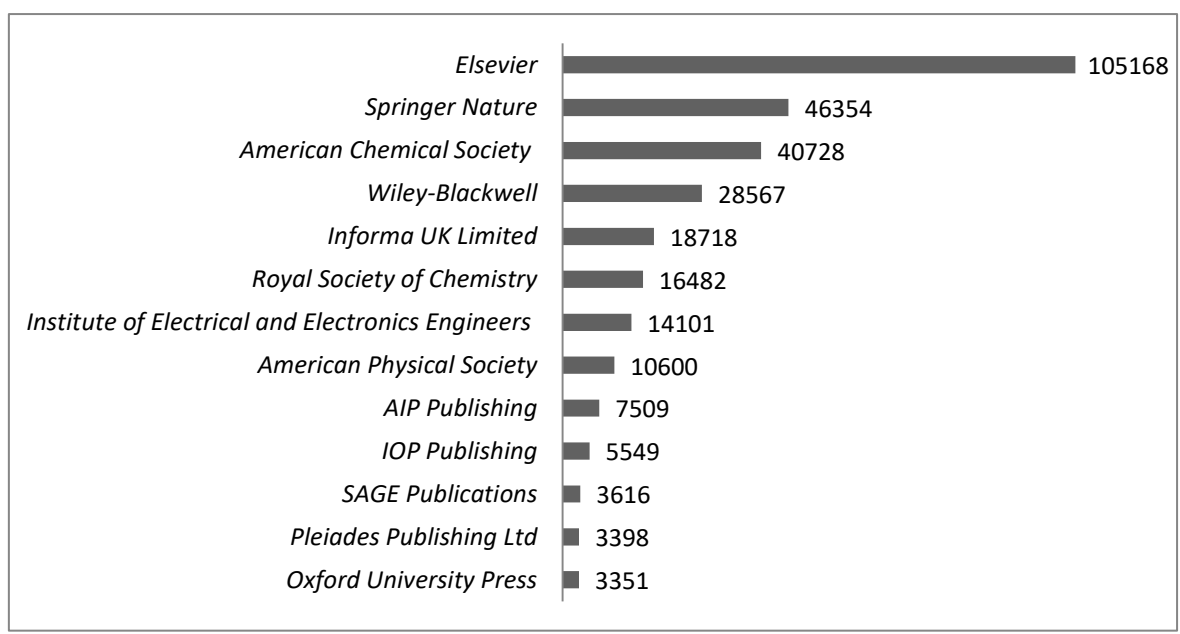

*Statistics of downloads from Sci-Hub by publishers (the publishers with downloads over 3000 are shown)*

The most documents downloaded from Sci-Hub by the Ukrainian users relate to Physical Sciences, in particular, chemistry, physics, and astronomy (69%). The second popular ones are Health Sciences 13% (medicine, pharmacology, and pharmaceutics) and Life Sciences 12% (neurobiology, biochemistry, genetics, and molecular biology). The least popular are Social Sciences 6% (transport, philosophy, political science). Correspondingly, the Ukrainian Sci-Hub users most frequently downloaded papers from the following journals: Journal of the American Chemical Society (6769 downloads), The Journal of Organic Chemistry (6038 downloads), Physical Review B (4325 downloads), Journal of Medicinal Chemistry (3712 downloads), and Tetrahedron Letters (3226 downloads).

If to compare the Sci-Hub download statistics of Ukrainian Internet users by fields of science with the data on effectiveness of their work in 2014—2015 based on Scopus (http://www.scimagojr.com/mapgen.php), one can see that the most publications of Ukrainian researchers relate to physics. The available data are not sufficient to state that when doing research the Ukrainian authors used mainly the documents downloaded from Sci-Hub, however, it is definitely that the researchers in the mentioned disciplines need a proper information provision and access to leading reference and full version databases.

This research does not consider the legal and ethical aspects of admissibility of downloads from illegal resource Sci-Hub or use of other resources and tools of Black Open Access. However, it should be noted that it is inadmissible to rely exclusively upon pirate web resources while developing and implementing the government strategy in the field of science. However, in Ukraine, there is no national subscription to information resources of leading publishers with the Ukrainian users very sporadically provided with legal access to them because of the lack of funding.

In addition, the researchers need a regular access not only to the latest results of research, but also to the archive documents. Even upon expiration of subscription period, employees from educational and R&D establishments may get open access to previously prepaid content. This possibility is guaranteed by LOCKSS (https://www.lockss.org) initiative of Stanford University that aims at developing and implementing legal tools for easy and cheap storage of electronic copies of prepaid documents. However, no attempts to implement LOCKSS or other similar tools in the operation of Ukrainian institutes on a wide scale have been made so far.

The detailed statistical data on the use of academic information resources at Ukrainian R&D and educational establishments, as a rule, are not disclosed, which makes it impossible to develop a qualitative policy for meeting information needs of Ukrainian researchers. Therefore, the data on downloads from Sci-Hub by Ukrainian researchers are especially important for understanding their information needs. The file containing the statistics of queries of documents in Sci-Hub by Ukrainian users is available at https://doi.org/10.6084/m9.figshare.5579092.v1, so everybody can use it for his/her own research.